# High-Resolution Scanning Tunneling Microscopy Imaging of Mesoscopic Graphene Sheets on an Insulating Surface

Elena Stolyarova[1], Kwang Taeg Rim[1], Sunmin Ryu[1], Janina Maultzsch[2], Philip Kim[2], Louis E. Brus[1], Tony F. Heinz[2], Mark S. Hybertsen[3], and George W. Flynn[1*]

[1] Department of Chemistry and Center for Electron Transport in Molecular Nanostructures, Columbia University, New York, NY 10027

[2] Department of Physics and Center for Electron Transport in Molecular Nanostructures, Columbia University, New York, NY, 10027

[3] Center for Functional Nanomaterials, Brookhaven National Laboratory, Upton, NY, 11973

*Corresponding author: gwf1@columbia.edu; Phone 212 854 4162; FAX 212 854 8336

**Abstract**

We present scanning tunneling microscopy (STM) images of single-layer graphene crystals examined under ultrahigh vacuum conditions. The samples, with lateral dimensions on the micron scale, were prepared on a silicon dioxide surface by direct exfoliation of single crystal graphite. The single-layer films were identified using Raman spectroscopy. Topographic images of single-layer samples display the honeycomb structure expected for the full hexagonal symmetry of an isolated graphene monolayer. The absence of observable defects in the STM images is indicative of the high quality of these films. Crystals comprised of a few layers of graphene were also examined. They exhibited dramatically different STM topography, displaying the reduced three-fold symmetry characteristic of the surface of bulk graphite.

## Introduction

Since the first reports of experiments on stand-alone, single-layer graphene crystals, (1) this remarkable 2-dimensional material has attracted great scientific interest (2-5). There are two alternative approaches for producing graphene layers. In the first method sample layers are mechanically exfoliated from bulk graphite crystals; in the second method a surface, such as silicon carbide, is "graphitized" under vacuum conditions. (6, 7) The strength of interaction between the underlying substrate and the graphene film is an issue of importance in the study of these materials of monolayer thickness. Very recent results using Angle-Resolved Photoemission Spectroscopy (ARPES) (7, 8) on single and few layer graphene samples has, for example, shown that interactions between a graphene film and a SiC substrate can be considered weak. On the other hand, several earlier STM studies of graphitized surfaces, such as Ir(1 1 1) (9), Pt(1 1 1) (10, 11), and SiC (6), have also been performed. In these experiments, the structure observed by STM was strongly influenced by the interaction between the graphitic layer and the underlying substrate, and features unambiguously associated with the electronic properties of an isolated graphene layer could not be identified. The purported differences in the strength of the graphene-substrate coupling may reflect different sample preparation methods and/or varying sensitivities of the STM and ARPES techniques to these interactions.

Here we present results of an STM study of single-layer graphene films prepared by mechanical exfoliation and probed on an insulating substrate. For these micron-sized samples, the STM topographic images show the hexagonally symmetric honeycomb structure expected for an ideal, unperturbed graphene crystal. STM images for multi-layer graphene films prepared in the same fashion display the reduced, three-fold symmetry characteristic of the surface of bulk graphite crystals. In addition to the local atomic-scale structure of single-layer graphene samples, we present measurements on the film's topography over the 100-nm length scale. Height variation on the order of 1 nm is observed. These investigations confirm the possibility of preparing high-quality graphene specimens, as atomically-resolved STM images of portions of this sample showed no evidence of defects or dislocations in the graphene crystallographic structure. We note that the samples were produced under ambient conditions and then subjected to microfabrication processing and exposure to various organic solvents. The fact that the films survived these severe conditions suggests that graphene holds promise not only for elegant scientific experiments, but also for novel electronic devices and sensors.

**Results and Discussion**

An optical image of the flake used in the experiments described below is shown in Figure 1 (a). A region with single-layer graphene (I) is visible on top of the wafer surface (III). Part of this flake (II) exhibits a higher optical density. We estimate this region to have a thickness of 5 graphene layers. The fabrication and identification of single- and multi-layer graphene samples by Raman scattering spectroscopy are described below. Figure 1(b) shows an optical image of the sample after electrode deposition around the flake. Most of the single-layer region (I) and a portion of the multi-layer region (II) are accessible for STM studies; the rest of the flake is buried beneath the gold film. Since the thin gold film is partially transparent, the contour of the entire flake of Figure 1(a) can still be seen.

Several hundred images were recorded for different positions of the STM tip. The graphene sample was found to be highly conductive so that tunneling occurs only between the STM tip and the graphene. Thus, the graphene itself and the gold electrode attached to the fringe of the flake (Figure 1 b) provide a return path for electrons in these experiments. In the measurements, the bias was set at +1V sample potential (STM images became unstable at low bias voltages), and a tunneling current of 1 nA was chosen. In the regions that were identified as consisting of single-layer graphene, a honeycomb structure was observed. Figure 2 (a) shows such an image over a 1 nm$^2$ area. No atomic defects were found in our images. This indicates the high quality of the graphene films produced by the present technique. For comparison an example of a 1 nm$^2$ image recorded on a multi-layer flake is shown in Figure 2 (b).

The characteristic features of the STM images of Figures 2 (a) and (b) are readily interpreted in terms of the A-B stacking of the graphene planes in graphite. In bulk graphite, the carbon atoms on the surface are not equivalent. Half of the carbon atoms in the surface layer are located above carbon atoms in the adjacent, lower layer (A-type atoms); the other half are sitting over a void (B-type). This asymmetry in the surface atom electronic environment results in a three-fold symmetry ("three-for-six") pattern in which three bright or dark features can be observed for each set of six carbon atoms, consistent with the structure shown in image 2 (b) (12, 13). This behavior is also present for graphene flakes that are two or more atomic layers thick.(14) For single-layer sheets of graphene this asymmetry is removed. Consequently all surface carbon atoms are identical, and a symmetrical honeycomb structure is observed in the STM image. Lattice defects, point-like interactions with the underlying substrate, or folding of the single-layer graphene sheet would be expected to cause significant perturbation of the local electron density and, thus, be reflected in the STM topography. None of these possible features were observed in the present study over the region of the sample investigated under atomically resolved STM.

It has been argued theoretically that perfectly flat, two-dimensional crystals are not stable (15). Indeed, a very recent TEM experimental study has demonstrated nanometer scale structural deformations in free-standing graphene films(16). In order to examine mesoscopic graphene structures as formed on an insulating substrate, the large scale topography of the present sample was investigated. A stereoscopic, large-area (100 nm×62 nm) STM image of a single-layer graphene sheet (from region I of Figure 1) is shown in Figure 3. This image has been corrected using plane height offset obtained from a second-degree Line Mean Square (LMS) fitting routine. In Figure 3, the vertical scale is enlarged to highlight the details of the graphene "landscape". While roughness beyond the atomic-level is obviously present, the characteristic fluctuation in height is relatively modest. The observed height variation of ~ 0.5 nm, which occurs on a lateral scale of ~ 10 nm, is comparable to that measured by AFM for similar graphene samples prior to microprocessing for the electrode deposition. The height variation is also comparable to that determined separately by AFM for the underlying silicon-dioxide surface. (online supporting materials of ref. 1) Consequently, the observed non-periodic roughness may arise simply from the graphene film following (at least partially) the features of the underlying silicon-dioxide surface. In addition, TEM studies of suspended graphene films have demonstrated that free-standing graphene films are corrugated on a mesoscopic scale, with out-of-plane deformations up to 1 nm (16). Of course, STM probing is accompanied by the application of elastic forces on the graphene sample. (12). Single-layer graphene sheets are especially susceptible to deformation resulting from such forces, which may also contribute to the observed large-scale topography.

**Conclusions**

Single-layer graphene flakes isolated on a silicon dioxide surface have been identified and distinguished from multi-layer flakes using Raman scattering. UHV STM images of both single-layer and multi-layer flakes have been observed. While multi-layer flakes exhibit STM images with three-fold symmetry typical of bulk graphite, single-layer graphene crystals display a symmetric honeycomb structure in which all the surface atoms contribute equally to the tunneling images. Our STM studies demonstrate that single-layer samples of graphene prepared by mechanical exfoliation exhibit a high degree of crystalline order and only sub-nanometer fluctuations in height on a longer lateral scale. Remarkably, these favorable material properties persist despite the relatively harsh processing conditions used in preparation of the samples. Further, the single-layer graphene crystals, as gauged by the atomically resolved STM topography, show no prominent signs of perturbation induced by interaction with the underlying substrate. These materials are thus well suited for advancing both the science and technology of highly 2-dimensional systems.

**Materials and Methods**

**Sample fabrication:**

The graphene films for this study were prepared by the mechanical exfoliation of bulk Kish graphite (Toshiba) according to the procedure of Ref. (1) The exfoliated graphitic flakes were deposited on the surface of a silicon wafer covered by a silicon-dioxide film of 300-nm thickness. Graphene flakes of monolayer thickness were initially selected from the vast majority of thicker ones by visual inspection with an optical microscope.

Electrical contact to the graphene was obtained by depositing gold electrodes around the flake using electron beam lithography. All non-conductive silicon dioxide regions of the surface were covered by a layer of gold because accidental positioning of the STM tip on an insulating region causes a tip crash and permanent tip damage. The typical procedure for micro-fabrication of this structure was performed by spin-casting a two-layer resist of MMA/MAA copolymer (first layer, Microchem) and 950K PMMA (second layer, Microchem). Resist around and at the edges of the flake was removed by electron-beam exposure with an FEI SEM, followed by development in MIBK: Isopropanol 1:3. A metal film (1-nm Cr/18-nm Au) was then deposited with an electron-beam evaporator (Semicore SC2000). The metal film present above the graphene flake was removed by lift-off of the underlying resist in acetone. The roughness of this resulting graphene film did not exceed 1.5 nm, as measured by atomic-force microscopy (AFM). The area of the gold electrode created by this method was ~500 $\mu m^2$. As a last step, a further layer of gold (150 nm) was deposited on uncovered areas of the wafer using a shadow mask. This film provided good mechanical and electrical contact between the graphene crystal and the STM sample holder.

**Raman spectroscopy:**

Recently it has been shown that Raman spectroscopy is a reliable, non-destructive tool for identification of single- and multi-layer graphene samples. (17, 18) Prior to the present STM studies, the graphene sample shown in Figure 1 b was characterized with Raman spectroscopy These Raman measurements were performed with a microscope set-up in the backscattering geometry, under ambient conditions at room temperature. The output of an Ar ion laser (457.9 nm) was focused to a spot size of ~1μm$^2$ to permit different spatial regions of the graphene sample to be probed. The laser spot could be positioned on the sample to an accuracy of a few hundred nm. The dependence of the Raman scattering on laser power was examined to ensure that the graphene flake was not heated or damaged by the incident laser beam. Figure 4 shows the Raman spectrum of single-layer graphene (region I, Fig. 1b) and of a few-layer flake (region II, Fig. 1b) in the spectral region of the graphite G and D* modes. Gold electrodes, located a few micrometers away from the region under study, do not contribute to the Raman signal. The D$^*$ mode is the second-order optical phonon mode near the K point in the graphene Brillouin zone; it is particularly strong because of its double-resonant enhancement in the Raman process. The change in the electronic structure in going from single-layer graphene to two or more layers is most easily identified in the D* mode, since this mode changes from a narrow and symmetric feature to one exhibiting an asymmetric line shape on the high-energy side.(17) This behavior is observed in the Raman spectra shown in Fig. 4, demonstrating that region I is indeed a single-layer flake, while region II consists of several graphene layers. The increase of the G-band intensity going from single layer to few-layer regions of the sample further supports this assignment (18). Raman spectral patterns were repeatable at different spots in each region of the sample.

**STM measurements**

Sample imaging was performed with a low-temperature STM (Omicron, MODEL LT-STM) at liquid nitrogen temperature. The measurements were performed under ultrahigh vacuum (UHV) at a background pressure below $1.5 \times 10^{-10}$ Torr. After the sample was moved into the UHV chamber, it was annealed at 280 °C for 6 hours to remove the resist residue and other contaminants. An etched tungsten tip, also annealed before imaging, was employed to obtain the STM topographs. The images were analyzed with a software package (Image Metrology, SPIP version 4.4.6.0). The images represent raw STM topographs and have not been modified, unless otherwise indicated. The algorithm that allows a flake to be found and regions within the flake to be identified is similar to the method described in our previous work. (14) Briefly, the STM tip is positioned close to the flake using a telescope. After an initial landing, the STM is moved step-by-step with respect to the sample, and the surface is imaged after each step. The gold-flake boundary could be readily found and served as a reliable marker of the STM tip position. Using the optical image of the flake as a map, an unambiguous correspondence was established between the local STM images and the macroscopic position of the STM tip.

**Acknowledgments:** The authors thank Etienne De Poortere, Kirill Bolotin, and Melinda Han for invaluable help in the microfabrication of the samples used in this study. This work was supported by the National Science Foundation through grant (CHE-03-52582 to G.W.F.), the MRI program (CHE-04-21191), and the NSEC Program (CHE-06-41523); by the New York State Office of Science, Technology, and Academic Research (NYSTAR); and by the U.S. Department of Energy (DE-AC02-98CH10886 to M.S.H. and DE-FG02-03ER15463 to T.F.H.).

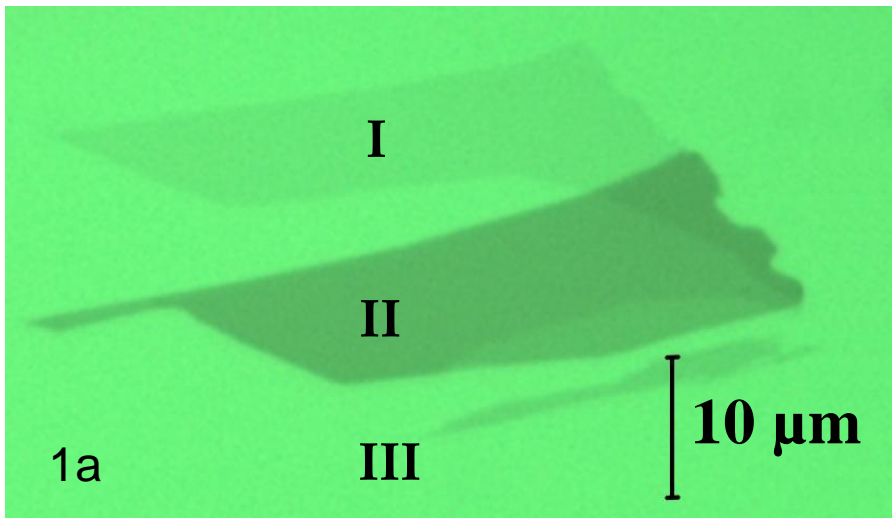

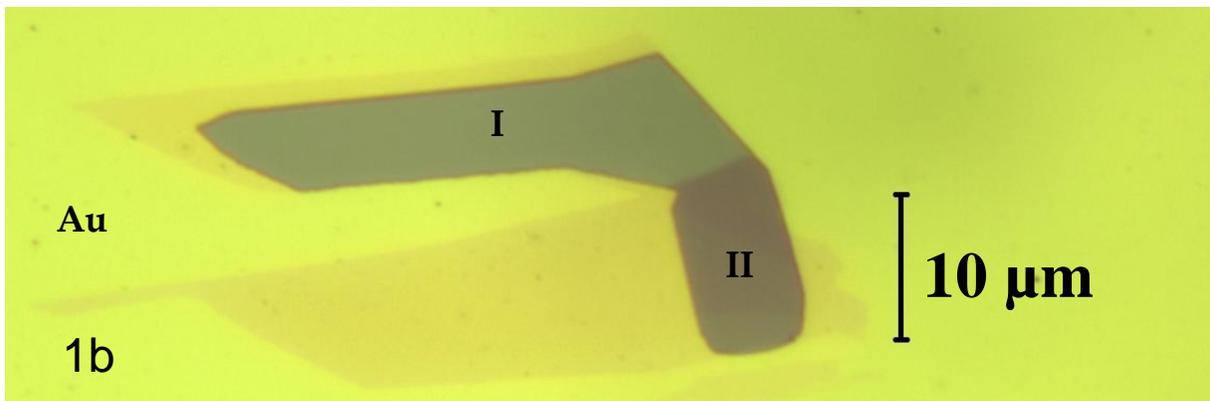

**Figure 1.** Optical microscopy images of the graphene flake examined in this study. (a) An image of the sample prior to deposition of the electrode. Three regions with different optical densities can be identified: I – single-layer graphene, II - multi-layer graphene, and III – the silicon-dioxide coated substrate. (b) An image of the same flake after the deposition of an 18-nm layer of gold. The gold electrode completely covers the substrate and partially covers the graphitic flake. The darker region is the uncovered part.

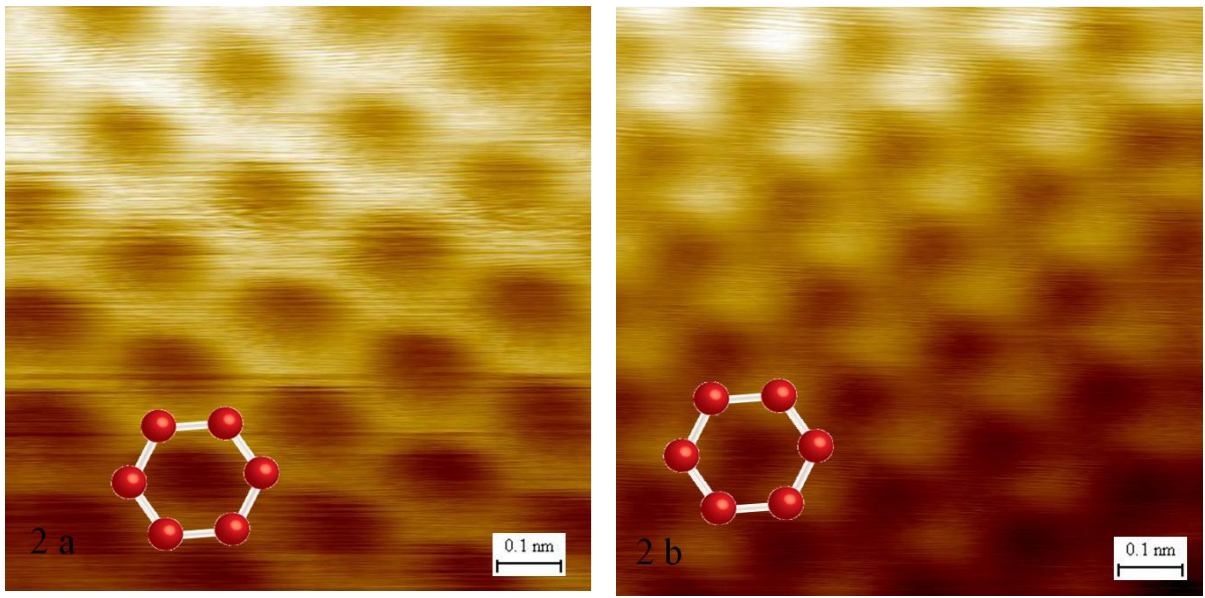

**Figure 2.** STM topographic images of different regions of the graphene flake of Fig. 1. The images were obtained with $V_{bias}= +1V$ (sample potential), I=1 nA, and a scan area of 1 $nm^2$. A model of the underlying atomic structure is shown as a guide to the eye. (a) Image from a single-layer of graphene (region I of Fig. 1). A honeycomb structure is observed. (b) Image of the multi-layer portion of the sample (region II of Fig 1). The characteristic "three-for-six" STM image of the surface of bulk graphite is observed.

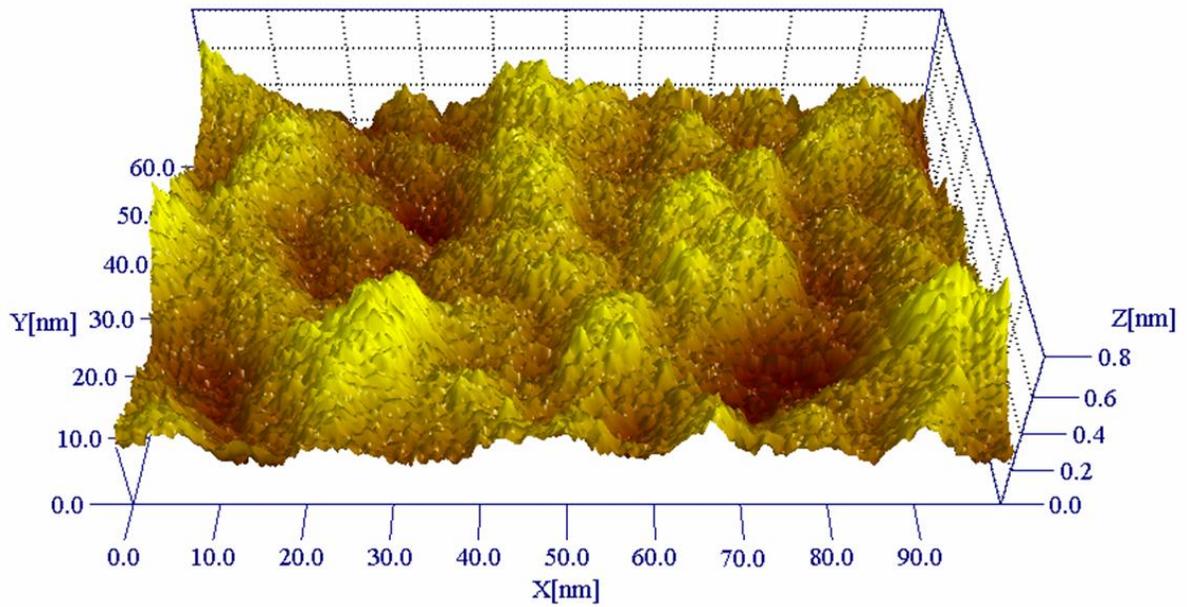

**Figure 3.** Stereographic plot of a large-scale (100×62 nm) STM image of a single-layer graphene film on the silicon dioxide surface. The STM scanning conditions were: $V_{bias}$= 1V (sample potential) and I=0.6 nA. The 0.8-nm scale of the vertical (Z) coordinate is greatly enlarged to accentuate the surface features.

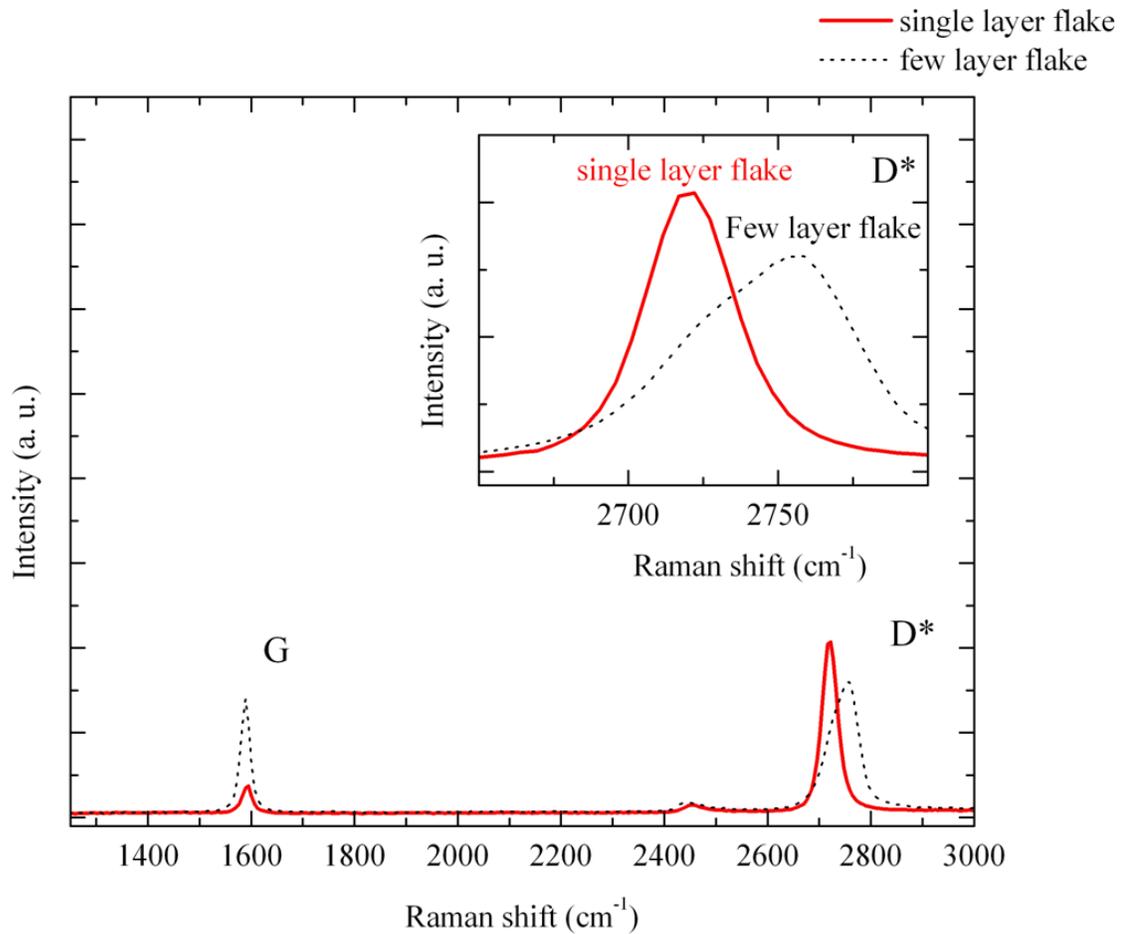

**Figure 4.** Comparison of Raman spectra at 457.5 nm for single-layer (solid line) and multi-layer (dashed line) regions of the graphitic flake described in Fig 1. The two intense features are the G peak at a Raman shift of ~1580 cm$^{-1}$ and the D* band at ~2710 cm$^{-1}$. The D$^*$ band (enlarged in inset) of a few-layer flake is blue shifted and broadened with respect to that of the single-layer graphene sample. Moreover, the D$^*$ peak of single-layer graphene is symmetric, while the D$^*$ band corresponding to the multi-layer sample has a complex asymmetric shape.